# The high strain-rate behaviour of three molecular weights of polyethylene examined with a magnesium alloy split-Hopkinson pressure bar.

F. Hughes*, A. Prudom, G. Swallowe

Department of Physics, Loughborough University, Loughborough, UK

*corresponding author: email f.t.r.hughes@lboro.ac.uk, telephone +44 (0)1509 228409

## 1. Abstract

A traditional split-Hopkinson pressure bar system has been modified by the addition of ZK60A magnesium alloy pressure bars in order to increase the resolution of data when examining specimens of low-density, high-density and ultra-high molecular weight polyethylene. It was found that the low density of the ZK60A allowed a decent increase in transmitted pulse amplitude, whilst its relatively high yield strength afforded long-term reliability of the system. The accuracy of data obtained from the fitted strain gauges was verified with the use of a high-speed video camera, and was found to be an excellent match.

### 1.1. Keywords

Hopkinson bar, Kolsky bar, high strain-rate, polyethylene.

## 2. Introduction

Polyethylene is perhaps the most commonly used polymer today, with applications ranging from the ubiquitous plastic shopping bag, through to skeletal joint replacements. It is easily produced, simple to mould, has good chemical resistance and is relatively strong. The mechanical behaviour of polyethylene in the quasi-static and low-rate strain regimes is well known. There are, however, many existing and potential applications for polyethylene for which it is necessary to understand how it behaves at higher rates of strain and accurate data at these strain rates is less commonly available.

The split-Hopkinson pressure bar is among the most commonly used experimental techniques for the determination of the dynamic behaviour of materials strained at rates of up to around $10^3$ $s^{-1}$.

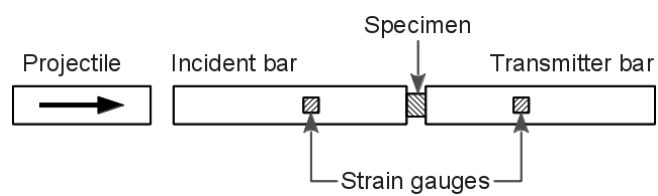

*Figure 1 - A simple SHPB system*

In use, a projectile or striker bar is fired at speed to collide with the incident bar, creating an incident strain pulse, $\varepsilon_I$, which propagates along the bar until it reaches the specimen. At this point, acoustic impedance mismatches between bar and specimen materials result in a portion of the pulse reflecting back along the incident bar, producing a strain $\varepsilon_R$, while some of the pulse is transmitted through the specimen and into the transmitter bar, with strain $\varepsilon_T$. It can be shown [1] that the engineering stress, $\sigma_S$ experienced by the specimen is

$$\sigma_S = E\left(\frac{A}{A_0}\right)\varepsilon_T \qquad (1)$$

where $E$ is the Young's modulus of the pressure bar material, $A$ is the cross-sectional area of the pressure bar, and $A_0$ is the initial cross-sectional area of the specimen. The engineering strain, $\varepsilon_S$ may be evaluated using the relationship

$$\varepsilon_S = \frac{2c_0}{\ell_0}\int_0^t \varepsilon_R \, dt \qquad (2)$$

where $c_0$ is the longitudinal wave velocity in the pressure bars and $\ell_0$ is the initial specimen length. Due to the energetic nature of the experiment, a typical SHPB system will be constructed using pressure bars made from a material with a high yield stress, such as high strength maraging steel. This allows a wide range of specimen materials to be forced to deform plastically while the pressure bars remain within their elastic limit. The traditional SHPB system is therefore an invaluable tool to study the behaviour of metallic solids. However, the need to understand low-impedance materials, such as the polymers examined in this work, has necessitated improvements to traditional SHPB systems in order to produce data of a suitably high resolution. Soft, low density specimens typically have a significantly lower acoustic impedance than that of the steel pressure bars, causing nearly all of $\varepsilon_I$ to reflect back along the incident bar. This leaves $\varepsilon_T$ with such a low amplitude that any data obtained will have a signal-to-noise ratio which is far too low to provide accurate, reliable data. A number of techniques are in common use as attempts to amend this problem.

## 2.1. Viscoelastic pressure bars

The use of viscoelastic pressure bars is a popular technique [2][3] employed to more closely match the impedances of specimen and pressure bar materials. Due to the viscoelastic nature of the bar material however, any wave input into the system will change as it travels along the pressure bars; the wave amplitude will attenuate, and the wave period will elongate. This makes it necessary to incorporate complex compensatory mathematical terms in the analysis of obtained data [4]. Viscoelastic pressure bars require complete homogeneity of the bar material, and could be unreliable in environments of inconstant temperature and humidity. Additional techniques have been developed to minimise some of the difficulties associated with the nature of the viscoelastic bars, including the use of velocity gauges to replace the strain gauges [5]. This technique, however, remains a complicated one.

## 2.2. Tubular pressure bars

Considering (2) above, it is clear that the amplitude of the transmitted wave is inversely proportional to the cross-sectional area of the pressure bar thus;

$$\varepsilon_T = \frac{\sigma_S A_0}{EA} \qquad (3)$$

hence tubular pressure bars may be used to reduce the overall cross-sectional area of the bars, increasing the amplitude of the transmitted wave [6]. Whilst this is desirable, it is necessary to have a flat surface in contact with the specimen, and so the bars must be 'capped' at each end. As may be seen in Figure 2, an end cap introduces an additional interface between the bars and the specimen, resulting in wave reflections within the pressure bar cap. This makes it more difficult to deduce exactly what the specimen is experiencing. As $\varepsilon_I$ is incident with the end cap, the energy will not be transferred uniformly into the cap, with the outer edge being loaded first, before propagating towards the centre. This can result in the non-uniform loading of the specimen, which may affect the reliability of any data obtained.

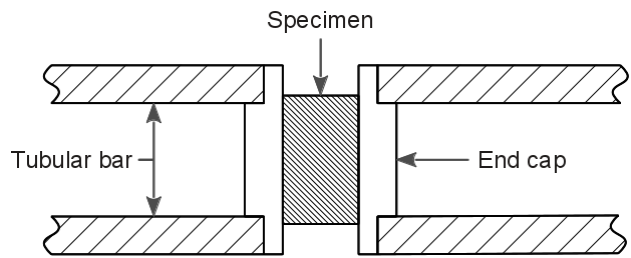

*Figure 2 – A tubular SHPB*

## 2.3. Low-density metallic pressure bars

In terms of providing reliable, easily interpreted data, the simplest SHPB design uses solid, metallic pressure bars. Due to the impedance mismatch in a traditional steel-barred SHPB system, it is desirable to minimise the mismatch with a view to increasing the amplitude of the transmitted wave. The acoustic impedance,

$$Z_0 = \rho c_0 \qquad (4)$$

is a function of the density, $\rho$, and the wave speed through the bar material, $c_0$. The use of low-impedance metallic pressure bars will therefore increase the amplitude of the transmitted signal. Previous studies have showed successful results from several low-density pressure bar materials including titanium [7] and alloys of magnesium [8]. In this study a number of suitable, low impedance bar materials were examined to test their accuracy and reliability. Specimens were examined in the SHPB and the results validated with the use of high-speed photography.

## 3. Materials

Discovered in 1933 by Gibson and Fawcett at ICI, polyethylene (PE) is one of the most widely used plastics today [9]. It is a polymerisation of ethylene, which has the formula $C_2H_4$, and consists of pairs of $CH_2$ groups, connected by a double bond as shown in Figure 3.

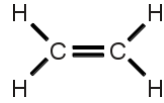

*Figure 3 – A molecule of ethylene*

Ethylene polymerises in contact with catalysts to produce long chains of $CH_2$ groups. Different lengths of these chains produce different classifications of polyethylene, each having unique mechanical properties. Longer chain lengths typically lead to an increase in the tensile strength of the polymer. The effect of longer chain lengths on the compressive strength is still a subject of debate, one which this work aims to further clarify.

Three different classes of polyethylene are examined here; linear low-density polyethylene (LLDPE), high-density polyethylene (HDPE) and ultra-high molecular weight polyethylene (UHMWPE).

## 4. Experimental

The SHPB system in use at Loughborough University was constructed using maraging steel pressure bars with an additional stainless steel 'pre-loading' bar [10] utilising a 20 cm long stainless steel projectile fired from a compact, evacuated gas gun [11]. Pressure bars and projectile all have a diameter of 12.7 mm. Cylindrical PE specimens measuring approximately 8 mm in diameter and 4 mm long were formed by compression moulding. Before making any modifications to the system, a benchmark was obtained by testing an LLDPE specimen in the maraging steel system.

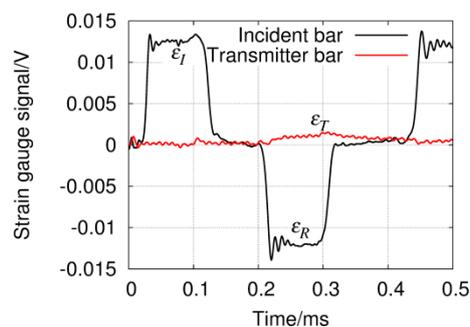

*Figure 4 – LLDPE specimen in a steel SHPB system.*

Figure 4 shows the strain gauge data collected from this experiment. The amplitude of $\varepsilon_T$ is very low with the percentage of the pulse transmitted being less than 10%. Figure 5 shows the stress-strain curve produced from this data. While it is possible to estimate an approximate yield stress for the material from this curve, the poor signal-to-noise ratio means that it could not be considered reliable. Any subtleties in the curve are completely lost in the background noise.

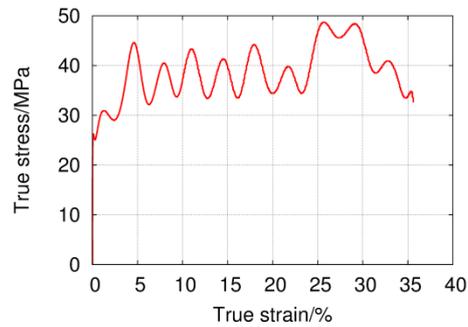

*Figure 5 – Stress-strain curve for LLDPE specimen in steel SHPB*

Because of the noise inherent in impact experiments it is common in Hopkinson bar work to use some data smoothing before analysis. Therefore, prior to analysis, this data, and all subsequent data presented in this work, was processed using a 7$^{th}$ order Butterworth low-pass filter with a cut-off frequency of 100 kHz.

Table 1 shows the properties of a range of materials suitable for use as pressure bars, along with LLDPE for comparison.

| Material | $\rho$ | $c_0$ | $\sigma_Y$ | $Z_0$ |
| --- | --- | --- | --- | --- |
| | kg m$^3$ | m s$^{-1}$ | MPa | kg m$^{-2}$ s$^{-2}$ |
| Aluminium | 2710 | 5119 | 80 | 13.9×10$^6$ |
| Aluminium, 7068 alloy | 2850 | 5061 | 550 | 14.4×10$^6$ |
| Magnesium | 1740 | 5029 | 95 | 8.8×10$^6$ |
| Nickel | 8900 | 4823 | 60 | 42.9×10$^6$ |
| Nickel, strong alloy | 8500 | 3597 | 1200 | 30.6×10$^6$ |
| Steel, stainless | 7930 | 5022 | 230 | 39.8×10$^6$ |
| Steel, maraging | 8100 | 5092 | 1800 | 41.2×10$^6$ |
| Titanium | 4540 | 5055 | 20 | 22.9×10$^6$ |
| LLDPE | 920 | 442 | 25 | 0.4×10$^6$ |

*Table 1 – The density ($\rho$), wave speed ($c_0$), yield stress ($\sigma_Y$) and acoustic impedance ($Z_0$) for a range of materials [12]*

Of all of these materials, magnesium has the lowest acoustic impedance, and so this was chosen as an initial pressure bar material. As expected, pure Mg bars provided a notable improvement in the amplitude of $\varepsilon_T$. The lower mass projectile, however, resulted in an $\varepsilon_I$ amplitude which was not great enough to deform the specimen significantly. The stress, $\sigma$, within the SHPB can be described [13] using

$$\sigma = \rho c_0 v \qquad (5)$$

where $\rho$ is the pressure bar density, $c_0$ the wave speed and $v$ the projectile impact velocity.

A measurement of the projectile velocity immediately prior to impact showed that it was travelling at approximately 10 m s$^{-1}$, and hence imparting a stress of around 88 MPa into the pressure bars. In order to increase specimen deformation to a more satisfactory level, it would be necessary to increase the projectile velocity, but since the induced stress at this projectile velocity is close to the yield stress of the Mg pressure bars, this would result in plastic deformation of the pressure bars. Notwithstanding, projectile velocity was increased to 20 m s$^{-1}$ ($\sigma$ = 176 MPa). The pressure bars did deform at this impact velocity, however specimen strain was improved such that the maximum strain during an experiment was 100%, which was deemed to be suitable.

A second attempt was made using the alloy ZK60A, which was chosen for its higher yield stress of 260 MPa. It has a chemical composition of 5.5% Zn, 0.5% Zr and 94% Mg. It has a density slightly greater than that of Mg ($\rho_{ZK60A}$ = 1830 kg m$^{-3}$) and a lower wave speed ($c_0$ = 4936 m s$^{-1}$) resulting in an acoustic impedance only marginally higher than pure Mg at 9.0×10$^6$ kg m$^{-2}$ s$^{-1}$. Pressure bars made from ZK60A were installed into the SHPB and a projectile was constructed with an increased length of 30 cm in order to extend the total strain duration. This was tested at projectile impact velocities of 20 m s$^{-1}$. The strain gauge data from an experiment with this configuration is shown in Figure 6. It can be seen that this is a considerable improvement upon the data illustrated in Figure 4.

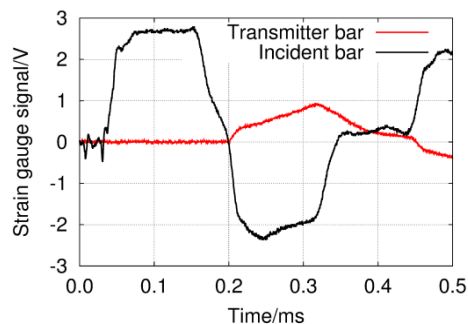

Figure 6 – LLDPE specimen in a ZK60A SHPB system.

The approximate doubling of the amplitude of the pulses on the incident bar may be attributable to the increase in projectile impact speed. The transmitted pulse amplitude has increased agreeably, with around 35% of the incident pulse being transferred into the transmitter bar, an improvement in resolution of greater than three times over the steel-based SHPB. It was decided that this configuration was more than acceptable for the comparison of the three PE materials. Full results are shown in the *Results and Discussion* section below.

From Equation (5), it would be possible to increase specimen strain further if desired. Given the relatively high yield stress of ZK60A, the projectile could be fired at a maximum impact velocity of 28 m s$^{-1}$, 40% faster than the tests performed here, without pushing the pressure bars past their elastic limit. This should be adequate to induce strains of 100% in most polymeric specimens.

It should also be noted that, during an experiment, constant strain rates were not achieved. Since the strains are relatively high, specimen heating must be significant, leading to the softening of the specimen during the experiment. This, of course, makes it impossible to give a precise strain rate for the experiment, and so all values quoted correspond to the strain rate at the point of maximum stress, as with the LLDPE specimen shown in Figure 7.

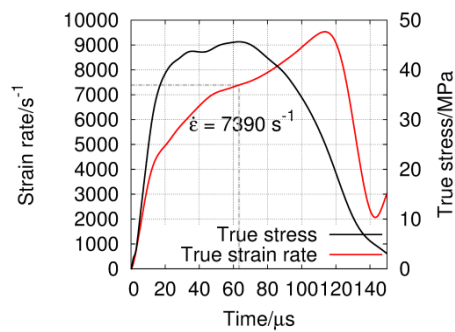

*Figure 7 – True strain rate and true stress vs. time curves for LLDPE specimen.*

## 5. Results and Discussion

Figure 8 shows the three stress-strain curves for LLDPE, HDPE and UHMWPE. The curves represent averages of three independent tests of each specimen, as per the data in [**14**]. The average strain rates at maximum stress for the data presented below were 7390s$^{-1}$, 7795s$^{-1}$ and 7386s$^{-1}$ respectively.

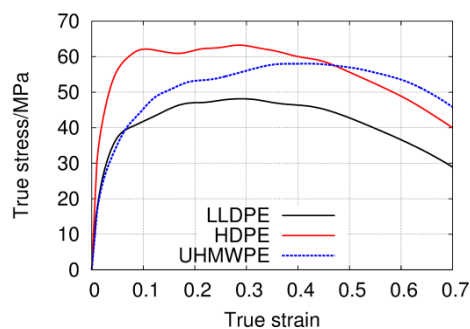

*Figure 8 – Stress-strain curves for LLDPE, HDPE and UHMWPE from a ZK60A SHPB*

The results in Figure 8 show reasonable agreement in shape with examples of high strain rate polyethylene properties in the literature, as for example the shape of the curves are similar to that reported by Walley and Field using a Steel Direct Impact Hopkinson Bar (DIHB) [**15**] and Hu *et al.* using an Aluminium Alloy SHPB [**16**] for HDPE, albeit not truncated as per the results in the referenced work. It is with this similarity, and given the quality of the results attained using the traditional steel Hopkinson bar setup typified by the data presented in Figure 5, that there is a reasonably high level of confidence that the data acquired by the Mg alloy bar system described above represents the material properties accurately.

From the presented results in Figure 8, the yield stress for HDPE is shown to be significantly higher than for the LLDPE and UHMWPE materials. It is also seen that UHMWPE experiences a large amount of strain hardening after yield not present in the other two materials, which would lead to greater energy absorption characteristics. This behaviour is different to that seen by the authors on the same materials tested at different strain rates using various apparatus [**14**].

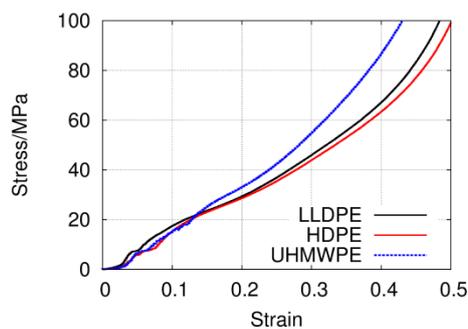

*Figure 9 - Stress-strain curves for LLDPE, HDPE and UHMWPE from a Universal Test Machine [14]*

Figure 9 shows the data for the same materials tested on a Hounsfield Universal Test Machine at strain rates (on average) of $3.05 \times 10^{-3} s^{-1}$, $3.19 \times 10^{-3} s^{-1}$ and $3.00 \times 10^{-3} s^{-1}$ respectively. The reason that the stress-strain curve gradient in the SHPB results presented in Figure 8 decreases shortly after yield, as opposed to rising as for the data reported above, is that in the SHPB experiment the high speeds - and therefore strain rates - involved in deforming the specimen cause a sharp rise in the temperature which leads to the decrease in the flow stress. Whereas the aforementioned Universal Test Machinery moves relatively slowly enough that any heat build-up caused by the deformation is dissipated and the test becomes essentially isothermal in nature.

Results obtained using an in-house designed Dropweight apparatus are shown in Figure 10.

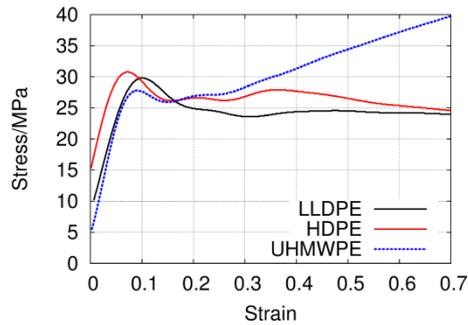

*Figure 10 - Stress-strain curves for LLDPE, HDPE and UHMWPE from a Dropweight apparatus [14]*

It is seen from these results, performed at an average strain rate of $1203s^{-1}$ and $1230s^{-1}$ for LLDPE and HDPE respectively, that initially all the curves rise steeply to yield in the region of 10% strain, dropping almost as sharply as they rose to a plateau before experiencing strain hardening at around 100% strain. For the case of the UHMWPE polymer, tested at an average strain rate of $823s^{-1}$, the strain hardening occurs almost immediately after yield and is much greater than for the other polymers.

## 5.1. Experimental Validation

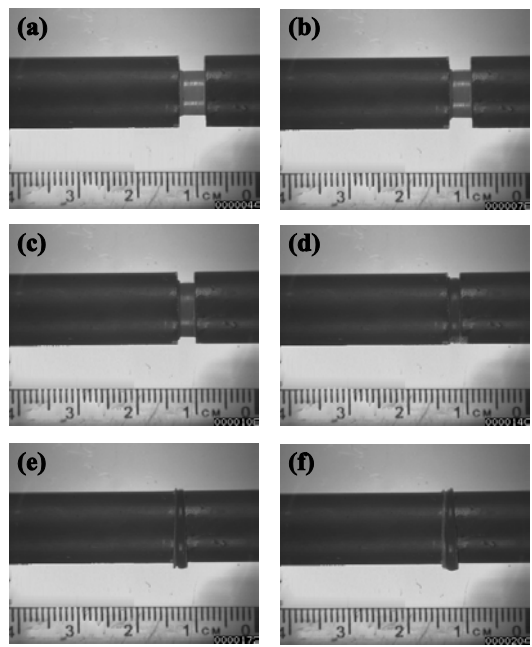

*Figure 11 - High-speed photography of LLDPE specimen in ZK60A SHPB. The images were taken at (a) 44μs, (b) 76μs, (c) 108μs, (d) 140μs, (e) 172μs and (f) 204μs.*

In order to validate that this SHPB configuration provided accurate, reliable data, a Shimadzu Scientific Instruments Inc. Hyper Vision HPV-1 high-speed digital video camera, on loan from the Engineering and Physical Sciences Research Council (EPSRC) Engineering Instrument Loan Pool, was employed to record the experiment. The camera was set-up perpendicular to the pressure bars such that the camera would record the change in the length of the specimen along the axis of impact as well as the change in specimen diameter at a 90° angle to the impact [**14**]. Video was recorded at a rate of 250,000 frames per second. Since the duration of the incident stress pulse is of the order of 150µs, this gives approximately 35 images showing specimen compression, each with a resolution of 312×260 pixels. Figure 11 shows six still images taken from one such recording at 32µs intervals. The incident bar may be seen on the right of the images, with the transmitter bar on the left

The images were used in ascertaining the conservation of volume of the specimen and to assess the dependability of the standard method of measurement capture, the bar-mounted strain gauges, at least until the specimen diameter becomes greater than that of the bar, such as shown in Figure 11, which will be discussed in Section *5.1.2*.

### 5.1.1. Volume Conservation

As mentioned briefly above, the images taken from the high speed camera were used in examining the conservation of volume within the experiment, and verifying the incompressibility of the specimen materials that is assumed throughout the main analysis. Measurements of the length and diameter of a specimen (HDPE for the following example) were taken by eye using a graphics program by enlarging the image to the pixel-scale and counting the pixels over the course of the bar impact, and the results are shown in Figure 12.

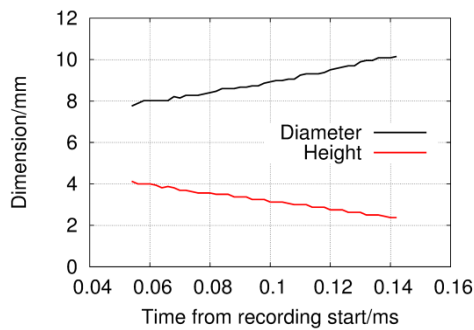

*Figure 12 - Dynamic height and diameter of specimen*

As can be seen in Figure 12, the height of the specimen, the lower line, decreases in a linear fashion as the specimen is compressed by the movement of the bar. It must be noted that the slight 'stepped' nature of the data is due to the measurement of the images to an accuracy of one pixel. Repeating the measurement for the specimen diameter shows that the diameter (red line) also increases linearly as expected.

Using these measurements and the usual formula relating the length and diameter dimensions of a cylinder to its volume, the dynamic volume of the specimen, compared to its measured initial volume during the test can be determined, as shown in Figure 13.

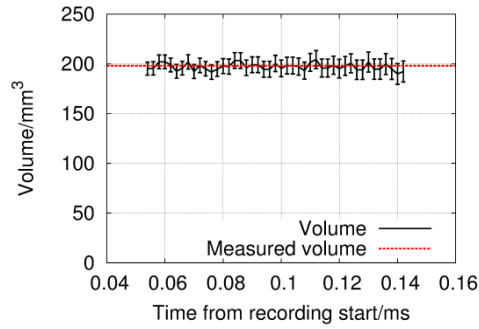

*Figure 13- Dynamic volume of specimen*

As is evident from the figure, given the uncertainty inherited from the measurement process, the volume is reasonably constant over the range of the figure which confirms the incompressibility of the material and substantiates the volume assumptions made in the main SHPB analysis. Also, as the volume is seen to stay constant and the material is essentially incompressible it can be said that Poisson's Ratio is approximately 0.5, confirming the value found in the literature [**17**].

### 5.1.2. Strain Gauge Signal Dependability

The use of the high speed camera equipment also gave an opportunity to confirm that the signals received from the strain gauges correspond to the actual material properties and validate the standard SHPB analysis technique for the new magnesium alloy pressure bars. In the following figures, measurement of the initial length, $\ell_0$, and momentary lengths, $\ell$, of the specimen, makes it possible to determine the instantaneous true strain, $\varepsilon_{\text{true}}$ using the relation

$$\varepsilon_{\text{true}} = \ln\left(\frac{\ell}{\ell_0}\right) \tag{6}$$

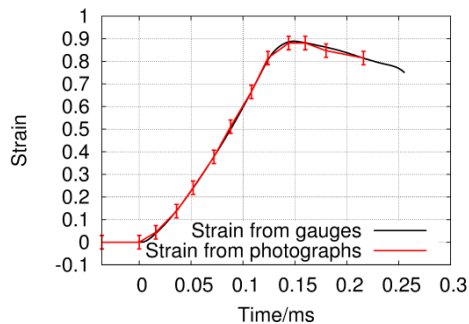

*Figure 14 - Photographic strain in HDPE specimen*

Figure 14 shows the result of the measurement of the photographic images for one of the HDPE specimens (chosen at random). As can be clearly seen, the strain measured from the images is very similar to the data recorded from the strain gauges, thereby validating the use of the strain gauges for accurate recording of the material strain without the time-consuming measurement of the photographic images.

As both stress and strain are measured using the same technique in the traditional SHPB experiment, it stands to reason that the stress measurement taken from the transmitter bar gauges is also very likely to be an accurate representation of what is experienced by the specimen. The uncertainty in the video data is due to the resolution of the images, with the specimen being measured to an accuracy of ± 1 pixel (corresponding to ± 0.14mm, or 3% of the initial specimen length).

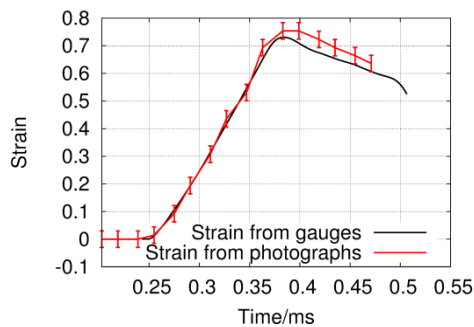

*Figure 15- Photographic strain in UHMWPE specimen*

Figure 15 is an example of the worst-fit from the tens of sets of analysed photographic images and relates to one of the UHMWPE specimens. Even though the strain gauge data lies outside of the aforementioned error margin for the unloading portion, it can be said that the data is still in reasonable agreement for the more important loading portion given the uncertainty in measuring the images by-eye.

As an attempt to see if the error could be reduced, a bespoke computer program was written to examine the photographic data, the results of which are shown in Figure 16 below. Using the program, it was not possible to measure specimen deformation beyond around 70% from the high-speed video since, as may be seen in the last two images in Figure 11, the specimen expanded outside the pressure bar area, so locating the ends of the bars was unreliable. This means that the strain gauge data is generally unreliable above approximately 70% strain.

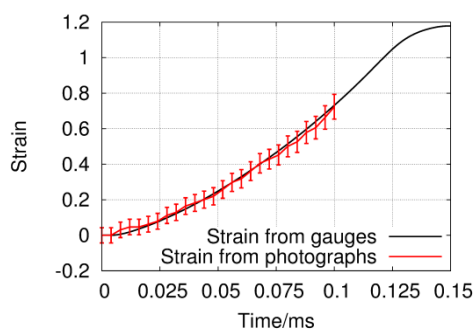

*Figure 16 - Photographic strain in LLDPE specimen*

### 5.2. Longevity

In previous studies [**18**] concern has been expressed that an SHPB system based upon Mg would not have enough strength to provide long term reliability. The SHPB system installed at Loughborough University has been in use for well over a year at the time of writing, and has performed many hundreds of experiments. To date it shows no obvious signs of wear. While it is unlikely that a ZK60A based system would have the very long term reliability of, say, a maraging steel system (primarily due to Mg and its alloys being more easily corroded than maraging steel), it certainly is strong enough to last a considerable time.

## 6. Conclusions

Modifying a traditional SHPB system with ZK60A pressure bars proves to be a significant improvement over the more common steel system when used for examining soft, polymeric specimens such as LLDPE, HDPE and UHMWPE. In the tests performed here, the ZK60A system proved to be capable of producing results with a resolution over three-times higher than a steel-based system. Total specimen strain achieved reliably in the PE specimens was up to 70% where the sample expands beyond the bar diameter, with the possibility of greater strains (or similar strains in less malleable specimen materials) achievable by further increasing projectile impact velocities. The improvements come with no obvious drawbacks or complications in data analysis.

For the most part, the materials tested behaved as expected; however, UHMWPE exhibiting a yield strength which is lower than that of HDPE was an unexpected result. This prompted an amount of experimental validation along with a comparison with alternative techniques, with both quasi-static and medium strain rates, which ensured that the results obtained are indeed accurate and reliable.